\documentstyle[twoside,epsf]{article}

\catcode`\@=11
\long\def\@makefntext#1{
\protect\noindent \hbox to 3.2pt {\hskip-.9pt  
$^{{\eightrm\@thefnmark}}$\hfil}#1\hfill}		

\def\@makefnmark{\hbox to 0pt{$^{\@thefnmark}$\hss}}	
	
\def\ps@myheadings{\let\@mkboth\@gobbletwo
\def\@oddhead{\hbox{}
\rightmark\hfil\eightrm\thepage}   
\def\@oddfoot{}\def\@evenhead{\eightrm\thepage\hfil
\leftmark\hbox{}}\def\@evenfoot{}
\def\sectionmark##1{}\def\subsectionmark##1{}}

\oddsidemargin=\evensidemargin
\newcounter{sectionc}\newcounter{subsectionc}\newcounter{subsubsectionc}
\renewcommand{\section}[1] {\vspace{12pt}\addtocounter{sectionc}{1} 
\setcounter{subsectionc}{0}\setcounter{subsubsectionc}{0}\noindent 
	{\tenbf\thesectionc. #1}\par\vspace{5pt}}
\renewcommand{\subsection}[1] {\vspace{12pt}\addtocounter{subsectionc}{1} 
\setcounter{subsubsectionc}{0}\noindent 
{\bf\thesectionc.\thesubsectionc. {\kern1pt \bfit #1}}\par\vspace{5pt}}
\renewcommand{\subsubsection}[1] {\vspace{12pt}\addtocounter{subsubsectionc}{1}
	\noindent{\tenrm\thesectionc.\thesubsectionc.\thesubsubsectionc.
	{\kern1pt \tenit #1}}\par\vspace{5pt}}
\newcommand{\nonumsection}[1] {\vspace{12pt}\noindent{\tenbf #1}
	\par\vspace{5pt}}

\newcounter{appendixc}
\newcounter{subappendixc}[appendixc]
\newcounter{subsubappendixc}[subappendixc]
\renewcommand{\thesubappendixc}{\Alph{appendixc}.\arabic{subappendixc}}
\renewcommand{\thesubsubappendixc}
	{\Alph{appendixc}.\arabic{subappendixc}.\arabic{subsubappendixc}}

\renewcommand{\appendix}[1] {\vspace{12pt}
        \refstepcounter{appendixc}
        \setcounter{figure}{0}
        \setcounter{table}{0}
        \setcounter{lemma}{0}
        \setcounter{theorem}{0}
        \setcounter{corollary}{0}
        \setcounter{definition}{0}
        \setcounter{equation}{0}
        \renewcommand{\thefigure}{\Alph{appendixc}.\arabic{figure}}
        \renewcommand{\thetable}{\Alph{appendixc}.\arabic{table}}
        \renewcommand{\theappendixc}{\Alph{appendixc}}
        \renewcommand{\thelemma}{\Alph{appendixc}.\arabic{lemma}}
        \renewcommand{\thetheorem}{\Alph{appendixc}.\arabic{theorem}}
        \renewcommand{\thedefinition}{\Alph{appendixc}.\arabic{definition}}
        \renewcommand{\thecorollary}{\Alph{appendixc}.\arabic{corollary}}
        \renewcommand{\theequation}{\Alph{appendixc}.\arabic{equation}}
        \noindent{\tenbf Appendix \theappendixc #1}\par\vspace{5pt}}
\newcommand{\subappendix}[1] {\vspace{12pt}
        \refstepcounter{subappendixc}
        \noindent{\bf Appendix \thesubappendixc. {\kern1pt \bfit #1}}
	\par\vspace{5pt}}
\newcommand{\subsubappendix}[1] {\vspace{12pt}
        \refstepcounter{subsubappendixc}
        \noindent{\rm Appendix \thesubsubappendixc. {\kern1pt \tenit #1}}
	\par\vspace{5pt}}

\newcommand{\textlineskip}{\baselineskip=13pt}
\newcommand{\smalllineskip}{\baselineskip=10pt}

\newcommand{\copyrightheading}[1]
	{\vspace*{-2.5cm}\smalllineskip{\flushleft
	{\footnotesize Quantum Information and Computation, Vol.~1, No.~0 (2001) 000--000 #1}\\
	{\footnotesize \copyright\kern2pt Rinton Press}\\
	 }}

\def\abstracts#1#2#3{{
	\centering{\begin{minipage}{4.5in}\footnotesize\baselineskip=10pt
	\parindent=0pt #1\par 
	\parindent=15pt #2\par
	\parindent=15pt #3
	\end{minipage}}\par}} 

\renewenvironment{thebibliography}[1]
        {\frenchspacing
	 \ninerm\baselineskip=11pt
         \begin{list}{\arabic{enumi}.}
        {\usecounter{enumi}\setlength{\parsep}{0pt}     
	 \setlength{\leftmargin 12.7pt}{\rightmargin 0pt}
         \setlength{\itemsep}{0pt} \settowidth
	{\labelwidth}{#1.}\sloppy}}{\end{list}}

\newcounter{itemlistc}
\newcounter{romanlistc}
\newcounter{alphlistc}
\newcounter{arabiclistc}

\newcommand{\fcaption}[1]{
        \refstepcounter{figure}
        \setbox\@tempboxa = \hbox{\footnotesize Fig.~\thefigure. #1}
        \ifdim \wd\@tempboxa > 5in
           {\begin{center}
        \parbox{5in}{\footnotesize\smalllineskip Fig.~\thefigure. #1}
            \end{center}}
        \else
             {\begin{center}
             {\footnotesize Fig.~\thefigure. #1}
              \end{center}}
        \fi}

\newcommand{\tcaption}[1]{
        \refstepcounter{table}
        \setbox\@tempboxa = \hbox{\footnotesize Table~\thetable. #1}
        \ifdim \wd\@tempboxa > 5in
           {\begin{center}
        \parbox{5in}{\footnotesize\smalllineskip Table~\thetable. #1}
            \end{center}}
        \else
             {\begin{center}
             {\footnotesize Table~\thetable. #1}
              \end{center}}
        \fi}

\def\@citex[#1]#2{\if@filesw\immediate\write\@auxout
	{\string\citation{#2}}\fi
\def\@citea{}\@cite{\@for\@citeb:=#2\do
	{\@citea\def\@citea{,}\@ifundefined
	{b@\@citeb}{{\bf ?}\@warning
	{Citation `\@citeb' on page \thepage \space undefined}}
	{\csname b@\@citeb\endcsname}}}{#1}}

\newif\if@cghi
\def\cite{\@cghitrue\@ifnextchar [{\@tempswatrue
	\@citex}{\@tempswafalse\@citex[]}}
\def\citelow{\@cghifalse\@ifnextchar [{\@tempswatrue
	\@citex}{\@tempswafalse\@citex[]}}
\def\@cite#1#2{{$\null^{#1}$\if@tempswa\typeout
	{IJCGA warning: optional citation argument 
	ignored: `#2'} \fi}}

\def\pmb#1{\setbox0=\hbox{#1}
	\kern-.025em\copy0\kern-\wd0
	\kern.05em\copy0\kern-\wd0
	\kern-.025em\raise.0433em\box0}

\def\fnt#1#2{\footnotetext{\kern-.3em
	{$^{\mbox{\scriptsize #1}}$}{#2}}}

\def\fpage#1{\begingroup
\voffset=.3in
\thispagestyle{empty}\begin{table}[b]\centerline{\footnotesize #1}
	\end{table}\endgroup}

\def\runninghead#1#2{\pagestyle{myheadings}
\markboth{{\protect\footnotesize\it{\quad #1}}\hfill}
{\hfill{\protect\footnotesize\it{#2\quad}}}}
\font\tenrm=cmr10
\font\tenit=cmti10 
\font\tenbf=cmbx10
\font\bfit=cmbxti10 at 10pt
\font\ninerm=cmr9

\font\eightrm=cmr8

\def\FigName{figure}%
\newbox\captionbox
\long\def\@makecaption#1#2{%
  \ifx\FigName\@captype
    \vskip\abovecaptionskip
    \setbox\tempbox\hbox{{\figurecaptionfont #1\hskip1em #2}}
	\ifdim\wd\tempbox< 28pc
	\centerline{\box\tempbox}
	\else
	{\figurecaptionfont #1\hskip1em #2\par}
\fi\else
  	\setbox\tempbox\hbox{{\tablecaptionfont #1\hskip1em #2}}
 	\ifdim\wd\tempbox< 28pc 
	\centerline{\box\tempbox}
	\else
	{\tablecaptionfont #1\hskip1em #2\par}%
	\fi   
 \vskip\belowcaptionskip
 \fi}
\InputIfFileExists{psfig.sty}{\typeout{^^Jpsfig.sty inputed...ok}}{\typeout{^^JWarning: psfig.sty could be be found.^^J}}
\InputIfFileExists{epsfsafe.tex}{\typeout{^^Jepsfsafe.tex inputed...ok}}
			{\typeout{^^JWarning: epsfsafe.tex could not be found.^^J}}
\InputIfFileExists{epsfig.sty}{\typeout{^^Jepsfig.sty inputed...ok}}{\typeout{^^JWarning: epsfig.sty could not be found.^^J}}
\InputIfFileExists{epsf.sty}{\typeout{^^Jepsf.sty inputed...ok}}{\typeout{^^JWarning: epsf.sty could not be found.^^J}}%
\def\fps@figure{tbp}
\def\ftype@figure{1}
\def\ext@figure{lof}
\def\fnum@figure{Fig.\ \thefigure}
\def\qed{\hbox{${\vcenter{\vbox{	          
   \hrule height 0.4pt\hbox{\vrule width 0.4pt height 6pt
   \kern5pt\vrule width 0.4pt}\hrule height 0.4pt}}}$}}

\def\one{{\mathchoice {\rm 1\mskip-4mu l} {\rm 1\mskip-4mu l} {\rm
1\mskip-4.5mu l} {\rm 1\mskip-5mu l}}}

\begin{document}
\setlength{\textheight}{7.7truein}    

\runninghead{NMR QIP and Entanglement} 
            { R. Laflamme {\it et al.} }

\normalsize\textlineskip
\thispagestyle{empty}
\setcounter{page}{1}

\copyrightheading{}	

\vspace*{0.88truein}

\fpage{1}
\centerline{\bf
NMR Quantum Information Processing and Entanglement}
\vspace*{0.37truein}
\centerline{\footnotesize 
Raymond Laflamme\footnote{laflamme@sciborg.uwaterloo.ca}}
\vspace*{0.015truein}
\centerline{\footnotesize\it Department of Physics, University of Waterloo, }
\baselineskip=10pt
\centerline{\footnotesize\it Waterloo, ON Canada, N2L 3G1.}
\centerline{\footnotesize\it Perimeter Institute for Theoretical Physics, }
\baselineskip=10pt
\centerline{\footnotesize\it 35 King 
Street N., Waterloo, Ontario, N2J 2W9, Canada.}
\centerline{\footnotesize\it Los Alamos National Laboratory,  }
\baselineskip=10pt
\centerline{\footnotesize\it Los Alamos, NM 87545, USA.}

\vspace*{10pt}
\centerline{\footnotesize David G. Cory\footnote{dcory@mit.edu}}
\vspace*{0.015truein}
\centerline{\footnotesize\it Department of Nuclear Engineering, }
\centerline{\footnotesize\it Massachusetts Institute of Technology, }
\baselineskip=10pt
\centerline{\footnotesize\it  Cambridge, MA 02139, USA }
\vspace*{10pt}
\centerline{\footnotesize Camille 
Negrevergne\footnote{camille@t6-serv.lanl.gov}, 
Lorenza Viola\footnote{lviola@lanl.gov}}
\vspace*{0.015truein}
\centerline{\footnotesize\it Los Alamos National Laboratory,  }
\baselineskip=10pt
\centerline{\footnotesize\it Los Alamos, NM 87545, USA.}

\vspace*{0.21truein}
\abstracts{
In this essay we discuss the issue of quantum information and recent 
nuclear magnetic resonance (NMR) experiments.
We explain why these experiments should be regarded as quantum 
information processing (QIP) despite the fact that, in present 
liquid state NMR experiments, no entanglement is found. 
We comment on how these experiments contribute to the future of QIP 
and include a brief discussion on the origin of the power of quantum 
computers. }{}{}



\vspace*{1pt}\textlineskip	
\section{General}	
\vspace*{-0.5pt}
\noindent
                               %

Can we implement QIP using liquid state NMR? 
At room temperature the nuclear spins of an ensemble of molecules in 
solution are in a highly mixed state. As the molecules are effectively
non-interacting and equivalent, the spin system is described by a density
matrix of the form
\begin{equation}
\rho =\frac{1}{2^{n}}e^{-\beta H}\approx \frac{1}{2^{n}}
(\one -\beta H+\dots )\:, 
\end{equation}
where $n$ is the number of distinguishable spins in each molecule 
({\it i.e.}, the effective number of qubits),  \( H \) is their 
Hamiltonian, and \( \beta =\hbar /kT \) the inverse thermal energy. 
The ratio of the energy of the system to the thermal one, $\beta H$, 
is on the order of  
\( 10^{-5} \).  This implies that the state is extremely mixed, 
and with the present number of spins (up to seven qubits), it is 
possible to show that such mixed state are 
separable \cite{zyczkowski:qc1998a,braunstein:qc1999a}. Thus, 
present day NMR experiments contains \emph{no} entanglement.  
And this could be the end of this paper (the shortest one we ever wrote).

It has been quoted in the literature that entanglement is the source
of the power of quantum computer \cite{ekert:qc1998b}.  If this would
be true, why then would people be interested in NMR quantum 
information technology at room temperature?  
In this paper we will comment on this question and suggest
some answers.  We will argue that, at least for small devices, 
entanglement is not the most important resource
for quantum computation (QC), and can even be absent.
Nevertheless, as we scale up to
large number of qubits, entanglement will probably become inevitable. 
We will explore some of the issues related to the apparent power of
QC and give arguments why liquid state NMR is an
interesting technology for QIP.  Two main reasons can be provided:
First, because the evolution of the spins is controlled by quantum 
mechanics; Second, because the claim that entanglement is a necessary 
condition for QC is only a claim.  As such, it should be questioned, 
and NMR technology is an interesting and experimentally accessible 
setting that challenges it. 
We should remind the reader that there is so far no proof that quantum
computers are more powerful than their classical counterparts except
in the so-called black box (or oracle) case {\it i.e.}, when a certain 
function is evaluated without a detailed circuit for its implementation, 
and the relevant complexity measure is the number of queries to the oracle.
An important example is Grover quantum search algorithm 
\cite{grover:qc1997a}.
While remaining somewhat unsatisfactory from the physical point of view, 
such black box models give strong indication that quantum computers are 
indeed more powerful, but not yet a proof.  
Even quantum algorithms attaining exponential speed-up, 
like Shor's algorithm for quantum factoring \cite{shor:qc1994a}, 
are only more efficient than the \textit{known} classical counterpart.  
This is important to keep in mind because the apparent extra power 
of quantum computers is so far only an assumption.  
The readers should not expect a theorem out of this paper, only 
ideas and directions for future research.

\section{Quantum information and entanglement}

To be clear we should precise that what we mean by QC
is the manipulation of quantum information in order to
solve a mathematical problem.  This manipulation is local,
as opposed to quantum communication which deals with information 
transmission between separate parties.  Unfortunately,
we do not have at present a unique definition of the concept of quantum
information itself.  In the community it is often associated with the 
presence of entanglement {\it i.e.}, it is a reasonable guess that when 
we have entanglement we have quantum information.   

In the case of quantum communication, where the parties involved can 
only share specially prepared states, perform local unitary 
transformations and classical communication, entanglement is an 
essential part of the protocol, and allows for an improvement of 
the channel capacities with respect to their classical values. 
Nevertheless, in QC everything is done locally and it is not clear 
whether entanglement is actually the essential ingredient to improve the 
efficiency of the computation.  As quantum computers seem more powerful
than their classical counterpart, we can ask the following questions:

Is it possible to have quantum information (hence do actual quantum 
computation!) without having entanglement? Is entanglement 
responsible for the apparent power of QC? 

Before we address these questions, let us summarize some of the
characteristics of entanglement.
Pure states are said to be entangled if they cannot be written as 
product states {\it i.e.}, 
\begin{equation}
|\Psi _{ent}\rangle \not =|\Psi _{1}\rangle \otimes |\Psi _{2}\rangle 
\:,
\end{equation}
for the simplest case of a bi-partite case.
For mixed states this simple definition fails. A mixed state is 
said to be entangled if it is non-separable, meaning that it cannot be 
be written as a convex sum of bi-partite states:
\begin{equation}
\label{rhosep}
\rho _{sep}=\sum _{i} a_{i}\left( \rho_1 \otimes \rho_2 \right)_i \:,
\end{equation}
where $\rho_1$, $\rho_2$ are density matrices for the two subsystems 
and the non-negative coefficients $a_i$ are interpreted as classical 
probabilities.  
There are interesting physical ensembles which are separable: 
for example, an equal mixture of all four Bell states can be
re-expressed as an equally weighted sum of the states \(
|00\rangle ,|10\rangle ,|01\rangle ,|11\rangle \), which are
separable.  This state remains separable even if there is a small
excess of, let's say, the state 
\( |\Psi _{+}\rangle=(|00\rangle +|11\rangle)/\sqrt{2} \),
such as 
\begin{equation}
\label{eqwerner}
\rho =\frac{1-\epsilon }{4}\one +\epsilon |\Psi _{+}\rangle 
\langle \Psi _{+}|\:,
\end{equation}
the so-called Werner state. 
If the parameter \( \epsilon \) is smaller than 1/3 this state is 
not entangled as it can be rewritten as a state of the form given 
in eq.(\ref{rhosep}). 

One of the quantitative definitions of the amount of entanglement 
present in two mixed qubits is related to the amount of pure EPR pairs 
that can be extracted from copies of the ensemble.  
This definition of entanglement
is based on concepts from quantum communication, not computation so
the following question appears natural: Is entanglement the right
notion to investigate the origin of the power of QC?

Entanglement is a new type of resource offered by quantum mechanics
which does not have a classical analog (states with entanglement
violate Bell inequalities, for example).  Although entangled states
cannot be used to communicate at distance, they can assist in some
communication tasks (for a review see \cite{cleve:qc1999b}).  Indeed,
it has been shown by Raz \cite{raz:qc1999a} that there is an
exponential gap in the amount of resources needed to solve a specific
problem if we use entanglement-assisted communication rather than 
classical communication.

Separable states do not contain the extra general correlations which 
appear in entangled states and are responsible for violating Bell 
inequalities. It is these correlations that are ultimately exploited 
in quantum communication.
Thus, highly mixed states describing room temperature liquid state NMR
(with small number of qubits) could not be used for quantum 
communication.

To summarize, quantum communication takes advantages of correlations 
which have no classical counterpart. On the other hand computation is 
inherently a dynamical process and the relevant resources could, at 
least in principle, be different.

It seems then reasonable to ask whether there is quantum information 
in states which are not entangled. There are a few examples of quantum
information without entanglement -- the work by Bennett {\it et
al.} \cite{bennett:qc1999a}, which demonstrates non-locality without
entanglement, and more recently the work by Ollivier and
Zurek \cite{ollivier:qc2001a}. The latter is based on the  
idea of comparing two different definitions of mutual information, 
obtained by generalizing the classical 
definition to quantum systems $X$ and $Y$:
\begin{equation}
\label{minfo1}
I(X,Y)=H(X)+H(Y)-H(X,Y) \:, 
\end{equation}
and 
\begin{equation}
\label{minfo2}
J(X,Y)=H(X)-H(X|Y)\:.
\end{equation}
In these equations, \(H(X,Y)\) is the joint entropy of the pair 
$(X,Y)$,  while \(H(X)\) and \(H(Y)\) denote the entropies for 
the reduced states of $X$, $Y$, respectively. \( H(X|Y) \) 
is the conditional entropy of the state of \( X \) given a 
projection of \(Y\) (summed over a complete set of projections of 
\(Y\)).
It turns out that \( I\) and \(J\) are equal for classical
systems \cite{cover:qc1991a}, or whenever a joint probability
distribution can be constructed for the individual systems 
\( X \) and \( Y \).
However, $I$ and $J$ differ otherwise. Quantum discord \( D \) is
defined as the difference \( D= J(X,Y) - I(X,Y) \).

An example of a state with non-vanishing discord is any states of the 
form (\ref{eqwerner}), provided that the measurement involved in 
(\ref{minfo2}) is a measurement of \( Y \) alone.  
Thus, even if these states may have no
entanglement, they have a non-zero discord {\it i.e.}, the information 
they contain is basically different from classical information.  
Those kind of states can be reached using generic evolutions in 
liquid state NMR.

There are other reasons to believe that, indeed, there is more to
QIP than just manipulating entangled states.  
It is an experimental fact that the behavior of nuclear spins
in a time scale shorter than the so-called \( T_{2} \) time is, even
at room temperature, surprisingly well described by a unitary 
evolution.  This \(T_{2}\) time corresponds to the decoherence time
where the phase of the state gets randomized. This means that  
although the system is in a highly mixed state, relative phases are 
preserved on that time scale. 
Moreover, we do not have an efficient way to simulate a
generic evolution of these systems.  So what is it in these evolutions
that makes it so hard to simulate?  For many years the NMR community
has attempted to find classical models to represent the behavior of
these high-temperature systems without succeeding \cite{braunstein}.  
This difficulty is well expressed in a claim by Ernst et al. 
\cite{ernst:qc1994a}: ``The dynamics of \emph{isolated} spins can be 
understood in terms of the motion of classical magnetization vectors.  
To describe coupled spins, however, it is necessary to have recourse 
to a quantum mechanical formalism where the state of the system is
expressed by a state function or, more generally, by a density operator.''  

This is not to say that there are no classical hidden variable theories
able to mimic quantum mechanics: one can always construct the hidden 
variable model which simulates the evolution of the system and arranges
the final state to reproduce exactly the quantum behavior.  This is 
possible as long as the systems is local ({\it i.e.}, we cannot rule
out classical communication between the qubits). However, this is 
true not only for highly mixed states but for pure states as well.  
In any event, these model always seem contrived to a high degree, and 
assume to some extent that the hidden variables are not counted as 
resources.  If indeed they are not taken into account, then a computer 
with hidden variables is as good as a quantum computer, although nobody 
believes that the hidden variables would really be resource-free.

This discussion emphasizes the fact that the distinction between
quantum and classical information processing is related to \emph{the
amount of resources} used, and not necessarily to the violation of
certain criteria such as the Bell inequalities.  In fact, it
is easy to show that allowing entanglement does not necessarily imply
that the information processing is hard to simulate classically.  An
example of this was given by Knill and Gottesman \cite{knill:qc1995a},
\cite{gottesman:qc1998} who realized that the
stabilizer operations (the operations which transform tensors product 
of Pauli operators into tensors product of Pauli operators) can
involve highly entangled states, but still states whose information 
processing capabilities are equivalent to classical ones.  
Yet, the most surprising fact about QIP is that it allows for certain
algorithms or protocols which are or seem exponentially more
efficient.

Once we have emphasized the distinction between quantum and classical
information processing through their efficiency properties, we should 
mention that liquid state NMR using the present methods for realizing
the required pseudo-pure states is not efficient.  Schulman and Vazirani
\cite{schulman:qc1998a} have shown that in the absence of noise there is 
an efficient algorithm for producing pseudo-pure states.  However, with
the level of noise and the amount of resources achievable by current
technologies, this method 
remains definitely impractical.

\section{Entanglement in NMR experiments}

As mentioned in the introduction, no entanglement has been
present in room temperature liquid state NMR experiments to date. 
Nevertheless, the degree of control available on the unitary evolution 
of the spin ensemble has allowed small algorithms to be demonstrated, 
clearly showing that quantum features can be implemented in this
sort of experiments.  The paper of \cite{braunstein:qc1999a} suggests
that states might exhibit entanglement at around thirteen qubits -- 
a number possible to reach, in principle, with the existing technology, 
although creating an entangled state would probably not be enough to 
motivate such experiment in practice.

However, there have been solid state NMR experiments where the state 
is likely to be entangled.  The first set is the investigation of
so-called multi-coherence or spin counting in calcium fluoride \cite{pines:qc1994a}.  
In this system, it is possible to control the evolution of an 
out-of-equilibrium state in such a way that after a desired time 
interval we can label various terms of the density matrix with a 
magnetic field gradient, and then time-reverse the Hamiltonian with 
an appropriate series of radio-frequency pulses.  The magnetization, 
which spreads to many nuclei in what the spectroscopists call
multi-coherence, can thus be reversed and brought back to its original
value with a signature of how large the achieved multi-coherence was. 
This is done in such a way that operators of the form \(
I_{+}^{1}I_{+}^{2}\dots I_{+}^{n} \)  (where \( I_{+} 
=(\sigma_{x}+i\sigma_{y})/2 \) for each spin) acquire a phase which 
evolves \( n \) times faster than the magnetization of a single spin.  
These states correspond to pseudo-pure cat-states.  
Coherences involving at least 60 spins (corresponding to a 
pseudo-pure cat state of 60 spins), are probably entangled -- although 
nobody has so far provided an analytic proof.

Another type of solid state experiment where entanglement is probably
present is the one investigating spin diffusion by Zhang and
Cory \cite{zhang:qc1998a}.  In these experiments, the spins of a 
calcium fluoride crystal are excited and left to interact (diffuse) with
their neighbors for some amount of time, and then the evolution is 
reversed.  If the magnetization is initially created locally, it will 
diffuse through the interaction with neighboring spins, and the relevant
question is, At which rate does this happen?  Attempts to determine the 
diffusion rate through classical simulations seem practically 
unfeasible, as around \( 10^{11} \) spins are involved in the quantum 
dynamics -- which would require \( \sim 2^{10^{11}} \) bits of classical 
memory.  With such a large amount of spins involved, the states are 
probably entangled (but again there is no proof of that yet).

\section{On the origin of the power of quantum computers}

There have been many suggestions for the origin of the power of quantum
computers: the size of the accessible Hilbert space, entanglement,
the existence of many universes, and the superposition principle
(or quantum interference). 

One of the first explanation for the power of quantum computation is
the size of the Hilbert space.  For example, for \( n \) spin 1/2
particles this size grows as \( 2^{n} \), thus it is tempting to claim
that there is an exponential gain compared to a classical system. 
But this is easily questionable since the state space of \( n \)
classical bits also contains \( 2^{n} \) distinct elements.  
One step further with this argument is to realize that the quantum 
system can be in a superposition state. Therefore, although the number 
of computational basis states is the same in the classical and quantum 
case, the \( 2^{n} \) 
computational states exhaust the possible configurations of a classical
system, whereas a generic quantum state is specified by an exponentially 
large set of complex amplitudes along the computational basis.
But again we could turn to a probabilistic classical computer 
(probably a better system to compare to the quantum case) and find that
there is an exponential number of states that can have non-zero probability
to occur.  On the other hand, there are definitely many quantum states 
which have no corresponding classical analog.

As discussed previously, the presence of entanglement can be invoked to 
explain the apparent speed up of QC.
Ekert and Josza \cite{ekert:qc1998b} argue
that if we start in a pure state and never produce an entangled state
we can follow the evolution of the system efficiently using a
classical description.  A set of classical tops can efficiently 
simulate a set of qubits which start in a pure state and
evolve without entanglement.  This can be seen from the Bloch sphere 
picture, where the state of a qubit is associated to a vector on the 
sphere. For many pure qubits which never get entangled the state is 
described as a set of Bloch sphere vectors.  These vectors can also 
be thought of as the directions of the angular momentum of spinning 
classical tops. Hence transformations that correspond to reaching
unentangled pure states can be mimicked by the classical system.
However, such a correspondence fails if entangled state are reached,
for entangled states cannot be described by a tensor product of Bloch 
spheres.  The above reasoning emphasizes the properties
of the \emph{state} during a computation.
However, we would like to also emphasize the amount of information 
necessary to follow the evolution of the system. In fact,  
an interesting point is to realize that the power of
quantum computers might take its origin in \emph{the properties of the
evolution as well as the states it uses}.  This argument seems also 
at first to give convincing evidence of the necessity of entanglement
at some stage of the computation -- but only if we consider pure states. 
Difficulties arise if we consider \emph{mixed} states.
In this case it is possible to investigate complex evolution operators 
which keep the state separable. For example, in the case of NMR, 
evolution operators ({\it e.g.} C-NOT gates) which would generate 
entanglement if applied to an initial pure state 
have been implemented, but because the state is highly mixed no entanglement
has been effectively created.  
Anybody who attempts to simulate these evolutions with
present algorithms, quickly realizes that they require an exponential
amount of resources as we scale up the simulation (as in the factoring
case, however, this does not exclude that we could find efficient
methods for such simulations in the future).  Another argument against
the point made in \cite{ekert:qc1998b} is the work by Bernstein and 
Vazirani\cite{bernstein:qc1993} and also Meyer on so-called
sophisticated quantum search without entanglement \cite{meyer:qc2000}.
He showed that in a black box model it is possible to have a quantum 
search algorithm that is more efficient than the classical case without 
requiring entanglement.

Most of the algorithms we know today have pure initial state, except
for possible quantum physics simulations.  Motivated by the goal of
investigating the power of NMR QC, two of the authors
studied the power of one bit of quantum
information \cite{knill:qc1998c}.  It is easy to verify that the
expectation value of the operator \( \sigma _{z} \) of the first qubit
in the circuit of Fig. 1 is given by 
\begin{equation} 
\langle \sigma_{z}^{1}\rangle = \mbox{Re}\, \mbox{Tr}\, [U\rho ]\:,
\end{equation}
where Re and Tr denote the real part and the trace operation, respectively.
Choosing \( \rho \) to be the identity density matrix gives us an
efficient way to find the trace of any unitary operator \( U \) which 
is efficiently implementable.  From a classical point of view this 
algorithm makes little sense, as most of the bits are in a completely 
random state. 
But for quantum systems, phases between states can be manipulated even
in the presence of highly mixed states as long as decoherence is 
negligible.  Calculating the trace of a generic operator  \( U \) is a
difficult task classically. 

\vspace*{2mm}

\begin{figure}[htbp]
\vspace*{13pt}
\centerline{\psfig{file=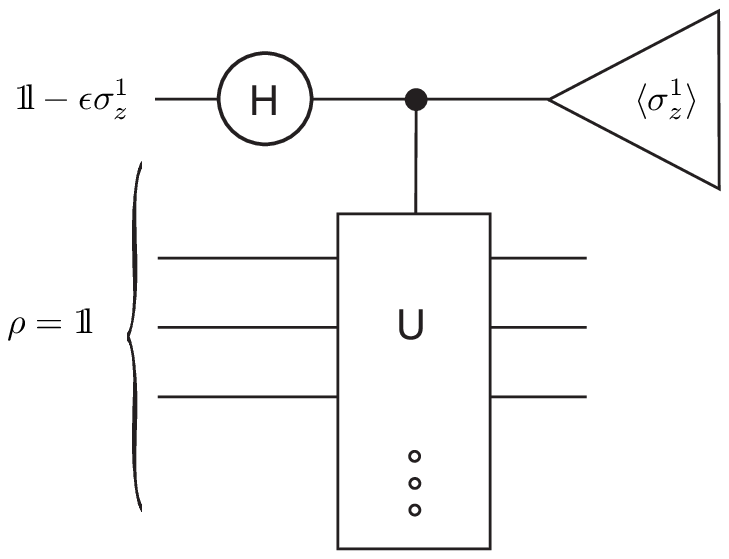}}  
\vspace*{13pt}
\fcaption{Quantum circuit evaluating the trace of the operator $U$ 
from an initial state of the form $\one -\epsilon\sigma_z^1$. $H$ 
represents the Hadamard gate on qubit 1. }
\label{fig1}
\end{figure}

The attractive feature of this computation is that the initial
state required for the quantum algorithm is scalable using the high
temperature limit of a thermal state.  The state is obtainable by
removing the polarization on all qubits except a single one.  Thus, in
the absence of noise, there is at least one (non black box) quantum
algorithm with no known efficient classical counterpart.  For thermal
states with little polarization, there will be no entanglement for a
small number of qubits.  For fixed polarization, entanglement will
probably appear for some threshold number of qubits, but there is no
indication to believe that at this number anything special occurs, or the
efficiency of the algorithm dramatically changes.  Therefore, this 
algorithm challenges the claim that entanglement is essential for speed-up.

An interesting point about the argument on entanglement is
that it implicitly relies on some notion of locality and an assumption
on how the physical system representing the qubits is behaving. 
After all we could think of having a single quantum system with \(
2^{n} \) dimensions and having the gates globally acting on these
states, thus mimicking operations on entangled states; why shouldn't 
physics allow this control?  In other words, the notion of entanglement is 
\emph{basis-dependent}, and the relevant question is how that preferred 
basis is chosen \cite{zanardi:qc2001}.

A third possibility is that the power of quantum computers originates in the
interference of the so-called multi-verse \cite{deutsch:qc1997}.  Even if 
some people find this suggestion compelling, it would be hard to prove or
disprove such hypothesis.

A final suggestion is that the essential ingredient is the superposition
principle of quantum mechanics.  The first consequence of this
principle is the size of the accessible Hilbert space; we already
discussed this point.  The second consequence is that the quantum
superpositions interfere.  At first sight it would appear that
classical wave mechanics also allows superposition and interference, 
so could we use water waves to quantum compute?  The answer seems to be 
No, because there are superpositions that have no classical analog for 
similar systems. 
However, there are mixed states which are separable but we do not
know how to transform into each other using local transformations and
classical communication.  We have described an example of such states 
previously, while discussing the notion of discord.

Maybe in the end we will learn that we need a combination of several of 
these resources.  Another possibility is that different resources are 
exploited for different approaches to obtain additional power from 
quantum mechanics.  Or possibility that it is impossible
to separate superposition, quantum dynamics, entanglement, projective
measurement for a scalable quantum device and that all these 
ingredients are important even though a dissection is necessary to 
arrive at this conclusion.  By
exploring concrete examples of QIP we hopefully will learn about the 
essence of quantum computation.
 
\newpage
\section{Conclusion}

In this essay we provided partial answers to the followings 
questions: 
\begin{quote}
$-$ Is entanglement required to have quantum information? \\
$-$ Is entanglement responsible for the apparent power of QC? \\
$-$ What is the origin of such power?
\end{quote}

We argued that there is more to QIP than just the creation and 
manipulation of entangled 
states, and commented upon the recent suggestion by Ollivier and Zurek 
to relate quantum information to discord.  We also described some solid 
state experiments where entanglement is probably present.
Finally, it is worth stressing that because we still do not know the 
exact origin of the power of QC, we do not know its necessary ingredients.  
However, it seems that the superposition principle and a state space 
dimensionality which grows exponentially should be part of the 
requirements.

Having said this, we believe it is important to remind the reader that 
the authors do understand the limitations of liquid state NMR
experiments, while at the same time they are convinced that these 
experiments provide a well-definite and accessible test bed for some of 
the ideas of QIP.  One achievement of the NMR
community has been to demonstrate control of generic unitary
transformations constructed from a set of universal gates (for example,
some of the work by Waugh and collaborators \cite{haeberlen:qc1968a}
was well before the interest in QC).  This is a
fundamental contribution to the field of QIP. Another fundamental 
contribution is the study of error mechanisms, which shows that the 
actually occurring errors seem to be reasonable enough to be compensated 
for by quantum error correction -- at least in the sense that the relevant 
error models are of the type expected, although the error magnitude is 
still too high for error-correcting and fault-tolerant methods to be 
fully effective. 
In spite of this, it has been possible to show that the unitary
control is able to take into account errors due to a variety of 
sources, such as mis-calibration, off-resonance,
and inhomogeneity, and that gate fidelities better than 0.999 can be 
currently achieved \cite{cory:qc2001a}.  
This also shows how important types of coherent errors 
can be controlled efficiently without having to recourse to full 
quantum error correction.

A further contribution is the knowledge acquired while translating
ideal quantum circuits into physically realizable ones.  As 
ideal circuits often assume some gates or operations which are not
directly available in a specific implementation, finding ways to 
circumvent these limitations is an important step forward.  
Moreover, the study of
scalability even for rather small numbers of qubits (two to seven) 
already indicates some of the difficulties encountered in going to larger 
numbers of qubits.  With two or three qubits, it is straightforward
to design a pulse sequence which implements a certain algorithm while
minimizing some of the imperfections of the device, but with five-seven 
it is much more challenging, and automation becomes  necessary.  
Understanding the optimal way to do this will be crucial for any device
which goes beyond a few qubits.

Finally, the present liquid state NMR experiments open the way for
a second generation of experiments: the solid state ones. Obviously,
this system will also have its limitations, but we hope it will allow
to better understand some of the difficulties of building QIP 
devices and to find ways to overcome them.

\nonumsection{Acknowledgments}
\noindent
The origin of this essay is a request by Dave Wineland (the editor)
to write a paper on entanglement and NMR. It is a pleasure to thank
Manny Knill for invaluable discussions and feedback on these 
subjects.  This work was supported by the Department of Energy under 
contract number W-7405-ENG-36, by the National Security Agency, 
the Advanced Research and Development Activity under
Army Research Office contract number DAAD19-01-1-0519, and
by the Defense Sciences Office of the Defense Advanced
Research Projects Agency under contract number
MDA972-01-1-0003.

\nonumsection{References}


\noindent
 
\begin{thebibliography}{000}

\bibitem{zyczkowski:qc1998a}
K.~Zyczkowski, P.~Horodecki, A.~Sanpera, and M.~Lewenstein.
\newblock Volume of the set of mixed entangled states.
\newblock {\em Phys. Rev. A}, 58:883--892, 1998.

\bibitem{braunstein:qc1999a}
S.~L. Braunstein, C.~M. Caves, R.~Jozsa, N.~Linden, S.~Popescu, and R.~Schack.
\newblock Separability of very noisy mixed states and implications for {NMR}
  quantum computing.
\newblock {\em Phys. Rev. Lett.}, 83:1054--1057, 1999.

\bibitem{ekert:qc1998b}
A.~Ekert and R.~Jozsa.
\newblock Quantum algorithms: Entanglement enhanced information processing.
\newblock {\em Phil. Trans. R. Soc. Lond. A}, pages 1769--1781, 1998.

\bibitem{grover:qc1997a}
L.K.~Grover.
\newblock Quantum mechanics helps in searching for a needle in a haystack.
\newblock {\em Phys. Rev. Lett.}, 79:325--328, 1997.

\bibitem{shor:qc1994a}
P.W.~Shor.
\newblock Algorithms for quantum computation: discrete logarithm and 
factoring. 
\newblock {\em Proceedings of the 35th Annual Symposium on Fundamentals
of Computer Science}, pages 124--134, 1994.

\bibitem{cleve:qc1999b}
R.~Cleve.
\newblock An introduction to quantum complexity theory.
\newblock quant-ph/9906111, 1999.

\bibitem{raz:qc1999a}
R.~Raz.
\newblock Exponential separation of quantum and classical
 communication complexity.
\newblock {\em Proceedings of the 31st ACM Symposium on Theory of Computing},
  pages 358--367, 1999.

\bibitem{bennett:qc1999a}
C.H.~Bennett, D.P.~DiVincenzo, C.A.~Fuchs, T.~Mor, E.~Rains, P.W.~Shor, 
J.A.~Smolin, and W.K.~Wootters.
\newblock Quantum nonlocality without entanglement.
\newblock {\em Phys. Rev. A}, 59:1153--1159, 1999.

\bibitem{ollivier:qc2001a} 
H.~Ollivier and W.H.~Zurek.
\newblock Introducing quantum discord.
\newblock quant-ph/0105072, 2001.

\bibitem{cover:qc1991a}
T.M.~Cover.
\newblock {\em Elements of Information Theory}.
\newblock Wiley, New York, NY, 1991.

\bibitem{braunstein}
Sam Braunstein in Les Houches in March 2001 commented
that this amounted to a proof by intimidation, and in some sense
it is true that it is not because a community has failed to
prove a proposition for many years that this is tantamount to
have proven the reverse. On the other hand, not mentioning that 
many people tried would be missing an important point.

\bibitem{ernst:qc1994a}
R.R.~Ernst, G.~Bodenhausen, and A.~Wokaun.
\newblock {\em Principles of Nuclear Magnetic Resonance in One and Two
  Dimensions}.
\newblock Oxford University Press, Oxford, 1994.

\bibitem{knill:qc1995a}
E.~Knill.
\newblock Approximation by quantum circuits
\newblock quant-ph/9508006, 1995.

\bibitem{gottesman:qc1998}
D.~Gottesman.
\newblock Theory of fault-tolerant quantum computation.
\newblock {\em Phys. Rev. A}, 57:127--137, 1998.

\bibitem{schulman:qc1998a}
L.J.~Schulman and U.~Vazirani.
\newblock Scalable {NMR} quantum computation.
\newblock In {\em Proceedings of the 31th Annual ACM Symposium on 
the Theory of Computation (STOC)}, pages 322--329, El Paso, Texas, 1998. 
ACM Press.

\bibitem{pines:qc1994a}
L.~Emsley and A.~Pines.
\newblock Lectures on pulsed {NMR}.
\newblock In B.~Maraviglia, editor, {\em Proceedings of the International
  School of Physics ``Enrico Fermi''}, volume CXXIII, pages 2949--2954, 1994.

\bibitem{zhang:qc1998a}
W.~Zhang and D.G.~Cory.
\newblock First direct measurement of the spin diffusion rate in a 
homogeneous solid.
\newblock {\em Phys. Rev. Lett.}, 80:1324--1327, 1998.

\bibitem{bernstein:qc1993}
E.~Bernstein and U. Vazirani.
\newblock {in \em  Proceedings of the 25th Annual ACM 
Symposium on the Theory of Computing, San Diego, 
1993 (ACM, New York, 1993), p. 11.}

\bibitem{meyer:qc2000}
D.A.~Meyer.
\newblock Sophisticated quantum search without entanglement.
\newblock {\em Phys. Rev. Lett.}, 85:2014--2017, 2000.

\bibitem{knill:qc1998c}
E.~Knill and R.~Laflamme.
\newblock On the power of one bit of quantum information.
\newblock {\em Phys. Rev. Lett.}, 81:5672--5675, 1998.

\bibitem{zanardi:qc2001}
P.~Zanardi.
\newblock Virtual quantum subsystems.
\newblock {\em Phys. Rev. Lett.}, 87:077901-1--4, 2001.

\bibitem{deutsch:qc1997}
D.~Deutsch.
\newblock {\em The Fabric of Reality}.
\newblock Penguin, New-York, 1997.

\bibitem{haeberlen:qc1968a}
U.~Haeberlen and J.S.~Waugh.
\newblock Coherent averaging effect in magnetic resonance.
\newblock {\em Phys. Rev.}, 175:453--467, 1968.

\bibitem{cory:qc2001a}
E.M.~Fortunato, M.A.~Pravia, N.~Boulant, G.~Teklemariam,
T.F.~Havel, and D.G.~Cory.
\newblock Design of strongly modulating pulses to implement 
precise effective Hamiltonians for quantum information processing.
\newblock Submitted to {\em J. Chem. Phys.}, 2001.

\end{thebibliography}
\end{document}